# The Unintended Carbon Consequences of Bitcoin Mining Bans: A Paradox in Environmental Policy

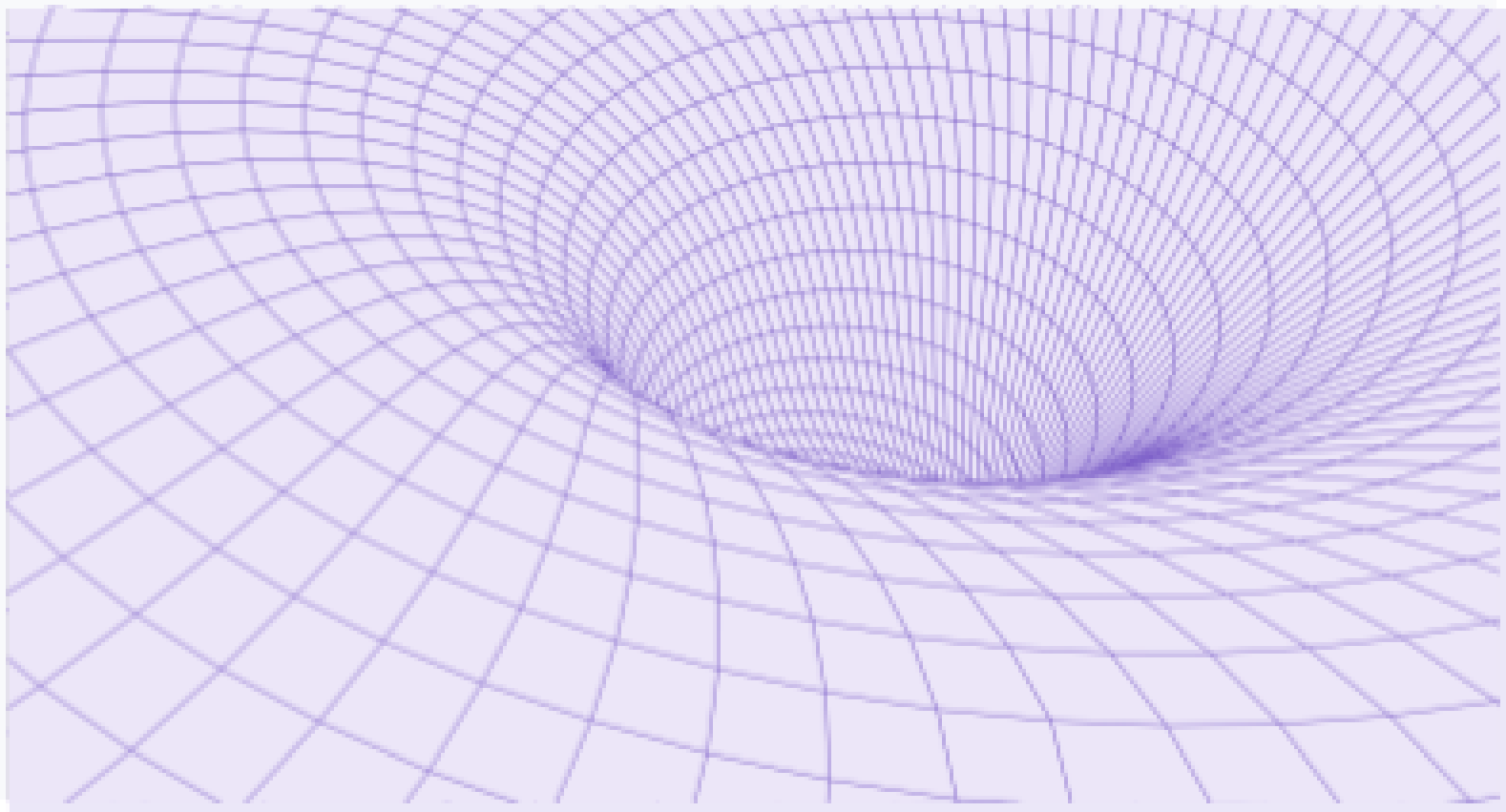

**Juan Ignacio Ibañez**
Exponential Science
University College London

**Paolo Tasca**
Exponential Science
University College London

**Aayush Ladda**
Exponential Science

**Logan Alred**
Exponential Science

# The Unintended Carbon Consequences of Bitcoin Mining Bans: A Paradox in Environmental Policy

Juan Ignacio Ibañez[123], Aayush Ladda[12], Paolo Tasca[123], Logan Aldred[2]
[1]Exponential Science, London, UK
[2]DLT Science Foundation, London, UK
[3]Centre for Blockchain Technologies, University College London, London, UK

***Abstract*—The environmental impact of Bitcoin mining has become a significant concern, prompting several governments to consider or implement bans on cryptocurrency mining. However, these well-intentioned policies may lead to unintended consequences, notably the redirection of mining activities to regions with higher carbon intensities. This study aims to quantify the environmental effectiveness of Bitcoin mining bans by estimating the resultant carbon emissions from displaced mining operations. Our findings indicate that, contrary to policy goals, Bitcoin mining bans in low-emission countries can result in a net increase in global carbon emissions, a form of aggravated carbon leakage. We further explore the policy implications of these results, suggesting that more nuanced approaches may be required to mitigate the environmental impact of cryptocurrency mining effectively. This research contributes to the broader discourse on sustainable cryptocurrency regulation and provides a data-driven foundation for evaluating the true environmental costs of Bitcoin regulatory policies.**

## I. Introduction

Mining is the key activity behind the consensus algorithm of Bitcoin: proof-of-work (PoW). Bitcoin's PoW relies on the generation of quintillions of hashes per second (work) by adversarial miners whose likelihood to collect block rewards is proportional to the share of the work that they contribute to the network. Notably, mining is an energy-intensive activity by design, which attaches a scarce resource to the ability to influence collective consensus, remediating the problem that digital identity creation is near-costless [1]. However, in recent years, this energy intensity has raised alarms around the world, posing concerns about the associated carbon footprint [2], [3].

### A. Literature Review

The technological intensity required to maintain cryptocurrency systems leads to high energy demand, e-waste, and carbon dioxide ($CO_2$) emissions [4]. This relationship between cryptocurrencies and the environment has been widely studied, with much focus on the need for environmental regulation to mitigate harm [4]. Despite existing research on crypto-market policy and mining bans, few studies explore the complex relationship between decentralized mining networks and regulation. In contrast, our study examines the effectiveness of regulation while considering the sustainability of the energy source and the likelihood of mining migration.

As cryptocurrency technology is relatively new, so are the regulations that attempt to control it, leading to a growing body of scholarship discussing the effectiveness and implementation of crypto-regulation. For instance, after China — one of the largest hubs for crypto-activity — implemented a series of cryptocurrency bans that forced the industry to relocate, some predicted that the industry would suffer as a result [5]. Contrary to these predictions, examination of the aftermath showed that industry connectivity actually increased after the ban [5]. While many studies focus on regulation's negative economics and operational impacts, they rarely address whether these bans cause unintended environmental consequences. Understanding these impacts is crucial to determine whether environmental crypto-mining bans truly achieve their goal of ecological protection.

Examining the energy sources powering crypto-mining is necessary to develop effective regulatory policies that mitigate environmental damage. As such, not all bitcoin mining is equal as countries utilize varying energy sources of varying environmental impact. For instance, a 2021 report ranked Ireland 9th overall in electricity use for bitcoin mining [4]. However, due to the country's high dependency on fossil fuels, it ranked sixth in cryptocurrency-related carbon footprints [4]. In contrast, Canada — which ranked 6th in electricity use for crypto-mining — benefits from its extensive use of nuclear and hydro-electric energy, resulting in a lesser environmental impact and placing the country at 9th in global carbon footprints from crypt-mining [4]. These findings, which align with our data, highlight that variations in crypto-mining will not have the same environmental impact across countries.

Further, comparative studies are commonly used in literature to investigate the global impacts of cryptocurrencies. Some studies compare countries based on their ecological footprints and environmental impact [6], while others examine regulatory approaches across nations, focusing on political and economic factors that may influence implementation [7], [8]. However, these concepts are rarely combined to account for the varying resources and regulatory strategies that lead to drastically different outcomes. In this study, we provide a more comprehensive view of the impact of crypto-mining by considering the decentralized nature of the Bitcoin blockchain and the complex contexts of different countries.

In response to a number of concerns, sometimes environmental, some regions around the world have introduced restrictive policy on crypto-mining, with China being a prominent example. In 2021, China imposed a ban on cryptocurrency mining, following earlier bans on exchanges and transactions [9]. The ban was driven by concerns that cryptocurrency could devalue the nation's fiat currency and that excessive resource use was contributing to environmental degradation [10], [11]. Yet, as a decentralized system, cryptocurrency networks are not limited to their localities, and therefore can move and act beyond physical or government borders. Rather than curbing global Bitcoin mining, to a great extent the ban merely displaced operations to nearby Kazakhstan, where regulatory environments were more lenient and electricity costs cheaper [9]. Since, there has been a 7.4% rise in electricity consumption in the first 9 months after China's ban, and Kazakhstan has become the second largest source of cryptocurrency mining globally [9]. This increase has resulted in the overuse of already strained resources and to multiple power-outages, putting businesses and citizens at risk [12].

Even in a case as prominent as China's landmark restrictive regulation, scholarly literature on the topic is exceedingly sparse. Currently, there is minimal literature published on the geopolitical interplay between cryptocurrency bans and the environment. As exemplified, the mining ban in China did not eliminate mining activity, and instead simply displaced it. Our study attempts to expand on this case by predicting how other crypto-mining bans, specifically on bitcoin, will impact the distribution of mining and therefore its energy usage.

Although literature addresses cryptocurrency's environmental impact, regulatory approach, and markets reaction, few studies combine these factors to get a holistic understanding of the complex environmental impact mining bans may create. Due to the dynamic and decentralized nature of cryptocurrency, regulation and industry decisions must be made with a comprehensive understanding of how mining and providers react to policy. This study aims to provide a comprehensive analysis of how bans on Bitcoin mining across countries can affect the global commons, by considering how mining redistribution across different energy contexts can lead to worsening environmental output.

### *B. The Effect of Mining Bans: Hash Rate Redistribution*

Although few articles in the field model ex-ante the effectiveness of mining bans, it is widely accepted that, in such cases, the hash rate would not be expected to permanently fall, but to relocate to other countries after a readjustment period [13], [14].

The reason for this effect is that Bitcoin mining is a zero-sum game: The amount of new bitcoin minted as coinbase transactions in every new block and the amount of transaction fees received by the block proposer do not depend on the amount of mining activity that is occurring at a given moment [16]. Across difficulty adjustments, if mining activity (hash rate) increases, block reward amounts do not increase correspondingly at the network level. Rather, competition between miners becomes fiercer and the expected rewards ofindividual miners fall. Conversely, if the hash rate decreases, the expected rewards of individual miners increase.

Naturally, this assumes everything else is held constant. In practice, under normal market conditions, other factors are not held constant when hash rate varies significantly over the long run. If the bitcoin price goes up, mining rewards increase in dollar terms, leading to an increased hash rate; the inverse effect happens if the coin's price goes down. This means that the variations in prices may be mitigated by the very variations in hash rate that they cause. If bitcoin's price increases, the increased profitability of mining will normally drive additional mining activity, and the increased hash rate will reduce expected rewards until marginal costs equal marginal revenue, i.e. driving profitability down until a new equilibrium is reached. If bitcoin's price falls, the decreased profitability of mining will expel some miners, again increasing expected revenue until a new equilibrium is reached.

While it is true that there are additional dynamics at play, such that e.g. short-term variations in hashrate cannot necessarily be explained by the model outlined above, this model is also widely accepted as explaining the fundamental aspects of the relationship between hash rate and the profitability of mining. This can also be applied to mining bans: If Bitcoin mining is banned in a particular jurisdiction, the hash rate can be expected to migrate to other jurisdictions without the ban due to fluctuations in profitability. The mechanism by which this geographic redistribution of hash rate operates is that, given that a miner's likelihood of receiving these rewards depends stochastically on the percentage of the global hash rate contributed by them to the network. If the global hash rate falls because a given region has banned mining with at least some degree of effectiveness, the rewards will stay the same at a network level, but the share represented by mining outside the region of the ban will increase, and so will the expected rewards from mining. This will attract new miners to the non-banned jurisdictions, until a new equilibrium is reached. If other factors are held constant, an equilibrium where marginal costs and revenues equalize can be expected at the same hash rate level as before. In other words, if the ban does not permanently impact bitcoin's price, the hash rate should eventually recover.

Certainly, it is conceivable that the hash rate struggles to recover because, in spite of an incentive for new miners to join the network, new mining equipment has not yet been manufactured. In this scenario, which can be aggravated by supply chain bottlenecks on key components such as silicon chips, hash rate may not instantly recover. Nonetheless, Bitcoin mining equipment is highly mobile; not just application-specific integrated circuit (ASIC)s but also mining containers, supplementary hardware, etc. This is known as location-agnosticity [16], and other factors such as the lack of a need for a grid connection and the low labor-intensity of the activity also

The same can be said of the block time, saving nuances related to difficulty adjustments [15].

Decreases in profitability may lead to other effects such as a more intense search for cheaper energy sources, especially in a context where capital expenses decrease through the commodification of mining equipment [16].

contribute to it. As a result, the increased profitability incentive in the other jurisdictions "attracts" the now idle ASICs of the ban jurisdiction to relocate to the former. In consequence, the hash rate can be expected to recover even without a very elastic supply curve for new ASICs.

### C. Existing approaches

The geographic redistribution aftereffect of a mining ban is widely accepted within the cryptocurrency community [14], [17]. However, it is unclear to what extent this is understood in policy circles.

Although there is an obvious shared intuition within the Bitcoin community that a mining ban in a low-carbon jurisdiction may lead to a redistribution of the hash rate to more carbon-intensive jurisdictions, to the knowledge of the authors, a quantification of this effect has only been attempted once, by Sang et al [14].

Sang et al predict the effect of a Bitcoin mining ban in China by comparing two methods. The first one, which we will call the "current proportion" method, assumes that after a mining ban, the hash rate is redistributed to the other countries in the same proportion as they currently have. The second one the authors name the "driving force" method, and assumes that the miners will relocate to the location where it would be most profitable to relocate, considering the following variables: Industrial energy prices, time of shipping and cost of shipping. Most of Sang et al's exploration is devoted to this second method.

Sang et al estimate that global emissions from a China ban would fall by 0.2% according to the current proportion method, but would rise by 4.4% following the driving force method.

### D. Our contribution

Sang et al's work is pioneering. However, it is only focused on China in a pre-ban context, it is based on 2021 data, and it is not Bitcoin-specific . This creates a gap in the literature that our article addresses.

Furthermore, we instead have a preference for the current proportion method, which remains underexplored by Sang et al. Sang et al's methododology encounters a number oflimitations. First, their implementation of the current proportion method only considers the top 8 countries by hash rate (China, the United States, Russia, Malaysia, Iran, Kazakhstan, Canada, Germany, and Ireland), while ignoring the rest. Second, they acknowledge a number of other factors that could influence the decision to relocate that they do not explicitly address, "such as local wars, terrorist attacks, theft, robbery, etc.". Third, there are also other very significant factors that they do not address nor acknowledge, but which could have an important influence in the miners' relocation decision, such as the regulatory environment, labor force qualification and availability, etc. Fourth, the driving force method may overestimate the influence of the costs and benefits from moving extant ASICs, and could underestimate miners' long-term focus. ASICs need to be replaced and new ASICs have to be purchased and shipped as a result. The decision on where to ship batches of ASICs can be regarded as more similar to an ex nihilo decision, not conditioned as much by the location of the replaced ASICs. With ASICs having a lifespan of two to five years [18], the economic lifespan being even shorter due to the entry to the market of more efficient machines, and the reasonable assumption that ASICs are on average halfway through this lifespan, the short-term considerations prioritized by the driving force method quickly wane. This is especially the case in a context in which yet another relocation would be admittedly costly. Fifth, the authors rely on industrial electricity prices, implicitly assuming that mining is happening on-grid. This overlooks the fact that a significant share of Bitcoin mining takes place off-grid, that the share off-grid mining is in a rising trajectory [16], [19]–[21], and hence that profit opportunities are often not dictated by country or province electricity price averages, but by highly local and specific price differentials (e.g. caused by variable renewable energy imbalances).

While the driving force method struggles to capture these factors, the current proportion method does not. In spite of risking present bias, the current proportion method prioritizes a market-based epistemology, under the assumption that market participants' revealed location decisions better approximate the landscape of opportunities in the market for hash rate relocation than what a limited model can predict. We select this method for our exploration, which is also easier to generalize to modeling on a global scale instead offocused on just one country.

The remainder of this paper is structured in the following way. First, we present the methodology of the paper. Second, we present our results. We then interpret these results and, finally, we conclude.

25 cryptocurrencies are in scope, a number of which have lost all relevance at present. This expanded the scope of the study back in time, but reduces its relevance at present, as well as prevents it from delving into the specificities of Bitcoin in more detail. This is also revealed in their reliance on secondary sources of miner location, which are admittedly based on a subset of mining pools's data on miners' IP addresses.

## II. Methodology

Our approach consists of a series of steps:

1) We create a map of global Bitcoin hash rate.
2) We calculate Bitcoin's energy consumption.
3) We identify the carbon intensity of all relevant jurisdictions. We calculate the pre-ban carbon footprint of Bitcoin.
4) Assuming the effect of a ban consists of redistributing the hash rate proportionally to all other jurisdictions, we calculate the post-ban carbon footprint of Bitcoin.
5) We introduce additional nuance by considering bans oflimited effectiveness, non-instantaneous adjustment periods, different carbon footprint types, and interjurisdictional coordination.

### A. The location of the Bitcoin hash rate

Establishing where Bitcoin miners are is crucial to understand their carbon footprint, as their location points to the type of energy they use and, thus, their carbon intensity. However, the location of the miners is not easy to establish for many reasons. Live maps of Bitcoin user nodes are available, but most user nodes are not miners and their locations can be completely unrelated. In addition, miners usually pool their resources in mining pools, and the IP location of the mining pool can also bear no relationship with the location of the actual miners. Finally, even when pool operators collect information on the miner's own IPs, miners often mask their true locations by resorting to Virtual Private Networks (VPNs).

At present, the most used database of Bitcoin miner locations belongs to the Cambridge Bitcoin Electricity Consumption Index (CBECI) by the Cambridge Centre for Alternative Finance (CCAF). The CCAF addresses many of these shortcomings by resorting to data validation techniques and performing ad hoc adjustments where required. Nevertheless, the CBECI has limitations. First, CBECI does not consider off-grid mining operations, whose carbon intensity may be completely unrelated to the carbon intensity of the grid of the jurisdiction in which the miners are. Notably, carbon-negative mining operations such as those based on otherwise vented or flared methane are excluded from CBECI. Second, CBECI miner location data is outdated at present, with the latest update corresponding to January 2022 [22].

An elaboration on some of the limitations of the CBECI model can be found in the justification for the Bitcoin Energy & Emissions Sustainability Tracker (BEEST) model of the Digital Assets Research Institute (DA-RI) [23]. We now explain how we address these limitations. Inspiration is explicitly taken from BEEST itself, but not the data, which is retrieved from Nodiens .

Our model starts with the CBECI miner map and updates it to the present day in the fashion of the BEEST model. Note that the CBECI map treats all countries as a single unit, except the United States and China, which are broken down into states and provinces, respectively. We consider the Kazakhstan and China ban by setting China's global hashrate share to 15% from February 2022 and reducing Kazakhstan's hashrate share to half ofits January 2022 level. The decreased hashrate from these two countries is redistributed proportionately to other countries based on their January 2022 shares. Then, we construct a database of known Bitcoin mining operations, on- and off-grid. We contrast the hash rate figures that we were able to map with CBECI data. Whenever our mapped hash rate exceeded CBECI's hash rate for that jurisdiction, our manually collected data was preferred and the rest-of-the-world (ROW) hash rate was reduced in proportion.

The methodology to identify known Bitcoin mining operations was to involve reviewing US Securities and Exchange Commission (SEC) filings of publicly listed companies, analyzing news articles, and monitoring updates on social media platforms like Twitter. We used the search terms "Bitcoin Methane Mining", "Bitcoin Renewable Mining", "Off-grid Bitcoin Mining", "Renewable Bitcoin Mining", "Flared Methane for Bitcoin Mining", "Vented Methane for Bitcoin Mining", "Sustainable Bitcoin Mining", "Green Bitcoin Mining", "Hydropower Bitcoin Mining", "Solar-Powered Bitcoin Mining", "Wind-Powered Bitcoin Mining", and "Geothermal Bitcoin Mining". Our findings were verified through company websites, company press releases and news articles from trusted sources to determine their official start dates by filtering searches based on known date ranges.

https://www.nodiens.com/

### B. *Bitcoin's energy consumption*

To calculate Bitcoin's energy consumption, we adapt the CBECI model. The key parameters used are network hashrate, block subsidy, transaction fees, mining equipment efficiency, electricity cost, and Power Usage Effectiveness (PUE).

First, we compute the operational expenses for all types of mining equipment identified by factoring in electricity costs and equipment efficiency. We then compare these expenses to a profitability threshold identical to CBECI's to determine which mining equipment is financially viable. Only those mining equipment deemed profitable are included in our energy consumption calculation.

Next, we establish the upper and lower bounds of energy consumption. The product of the hash rate and the energy intensity of per hash of the profitable mining equipment with the highest power usage provides the upper bound, representing the maximum possible energy consumption. In turn, the profitable equipment with the lowest power usage offers the lower bound, indicating the minimum energy consumption. Finally, we use the average efficiency of the mining equipment (i.e. an equally-weighted basket of profitable miners) to estimate the best-guess energy consumption for the Bitcoin network. This approach gives us a range of potential energy use, with the average providing our best estimate. In the remainder of this paper, this estimate, set at 16.91 GW (an annual energy demand of 148.12 TWh) is used.

### C. *Carbon intensity of jurisdictions*

Once established where the miners are and how much energy they are consuming, the next step is to calculate the carbon intensity of the energy usage. In this direction, we should distinguish on-grid from off-grid energy usage.

1) **On-grid energy usage**: To evaluate carbon emissions from various energy sources in global electricity production, we use the most recent data available from "Our World in Data" [24] for national statistics and from Ember Climate for data specific to U.S. states [25]. These datasets offer a detailed view of the electricity mix for countries and U.S. states. However, note that the data is current only through 2023 for most regions, with some countries providing data only up to 2022. To estimate the carbon emissions, we use the latest available percentage share of each energy source. For carbon intensity in Chinese provinces, we refer to 2023 data from the Chinese Academy of Environmental Planning [26].
2) **Off-grid energy usage**: For the miners identified as mining off-grid, we do not take the grid intensity of their jurisdiction, but calculate ourselves their carbon intensity based on the energy sources used by these miners. . We diverge from BEEST in this sense, as we do not filter out off-grid sustainable miners only and assume them to be zero emissions. Rather, we consider all miner types, and discriminate carbon intensities by energy source and according to different carbon intensity calculation types. See subsection II-E for an elaboration.

### D. Hash rate redistribution

At this point, we have established a map of Bitcoin miners around the world and their carbon footprint. We now model the scenario where the banning jurisdiction's hash rate goes to zero, and the subtracted hash rate is added to the ROW in proportion to the current share of each country. We calculate the carbon footprint before and after this redistribution.

### E. Introducing additional variables

While initially informative, some aspects of the model so far are not realistic. Hence we currently account for them.

*a) Bans oflimited effectiveness:* Bans are not fully effective at completely eradicating mining activity, as some ofit may be driven underground [31]. To account for this, we recalculate a limited effectiveness scenario based on the assumption that mining activity in the banning jurisdiction is reduced by 50% instead of 100%. This approach is taken from Nordhaus [32].

*b) Non-instantaneous adjustment periods:* Hash rate relocation is not instantaneous, hence during the adjustment period energy usage may actually be reduced, and the same will happen with the carbon footprint. To account for this, we assume a relocation period of two months, a conservative assumption which corresponds to the upper bound of the relocation period of Sang et al (55 days) [14]. While relocation should be progressive (e.g. we can assume that over a 2-month relocation period 25% of the hash rate is recovered in each of the fortnights), for simplicity we can assume an instantaneous recovery of the hash rate after a 1-month full suppression We calculate the carbon emissions of the banning country over one month. The difference between a month of pre and post-ban emissions approximates the one-off carbon avoidance from the frictional relocation effects.

We are unable to subtract these avoided emissions from our final results because the latter are annual emissions, and the one month of avoided emissions does not offer annualisable quantity but a one-off avoidance. Annualising it would require projecting future Bitcoin emissions and discounting them to their net present value, considering the evolving social cost of carbon throughout an unspecified period, which is beyond the scope of this article. For this reason, the one-off carbon avoidance caused by the ban's frictional relocation effects is presented separately in Appendix Table IV.

*c) Different types of carbon emissions:* Wind energy does not generate carbon emissions from the wind turbine itself. However, the manufacturing and other steps in the life cycle of a wind turbine do generate carbon emissions. This creates all sorts of problems, as one may justifiably ponder what fraction of the life cycle emissions of a wind turbine can be attributed to a Bitcoin miner that arrives after the windpark has been built and that had not been foreseen by the windpark owner as a possible energy buyer. Abstracting away from these questions, we simply consider both the LCA emissions and POG emissions of each energy source.

*d) Interjurisdictional coordination:* Considering the very realistic possibility of an EU ban on Bitcoin mining, we also model a scenario where all EU countries simultaneously ban Bitcoin mining, and the entire hashrate of the EU relocates to other countries.

## III. Results

The results of our simulation are presented in figures 1, 2 and 3, and in Appendix tables I to IV. The global map depicts the overall impact of Bitcoin mining bans on network carbon emissions across various countries. It highlights regions where a ban would lead to an increase in emissions versus those where it would result in a reduction. The map is color-coded to show the magnitude of emission changes, with darker colors indicating larger increases or decreases.

The US and China maps provides a detailed view of the emission impacts of mining bans at the state level. It demonstrates which states would see a reduction in emissions and which would experience an increase.

These results highlight the complex relationship between regional Bitcoin mining bans and their impact on the network's total carbon emissions. In some regions, banning Bitcoin mining may inadvertently lead to an important increase in emissions, while in others, it can significantly reduce them.

Notably, a mining ban in Canada would lead to the largest positive impact on emissions, increasing network emissions by approximately 5.6%, or 2.5 million tonnes of $CO_2$ annually. Similar trends are observed in countries where a significant portion of electricity is derived from renewable sources, or where off-grid mining is prevalent. For instance, a ban in Paraguay would increase POG emissions by 1.9M tonnes, while El Salvador (+650,000 tonnes), and Norway (+576,000 tonnes) would also experience increases in emissions. A complete ban in the European Union would increase POG emissions by 523,000 tonnes.

Conversely, countries with more carbon-intensive energy grids would indeed see a successful reduction in emissions from a Bitcoin mining ban, as intended. For example, a complete ban in Kazakhstan would reduce network emissions by approximately 7.6%, resulting in a decrease of 3.4 million tonnes of $CO_2$ annually. Other countries such as China (-2.1

The following point-of-generation (POG) carbon intensities are assumed, all in $CO_2$ eq g / kWh: 820 (coal), 490 (gas), 700 (other fossil), 0 (solar, wind, nuclear, hydro, other renewable) [27]. The following life cycle assessment (LCA) carbon intensities are assumed: 820 (coal), 490 (gas), 700 (other fossil), 48 (solar), 11 (wind), 12 (nuclear), 24 (hydro), 38 (other renewable) [27]. To calculate the carbon footprint of otherwise flared and vented methane, we assume 52.20 mega joules per ton of methane emitted [28], and that the following levels of carbon removal are achieved by each of the following methods: 0% (venting), 91.1% (flaring), 99.6% (mining) [29], [30]. This results in a negative carbon footprint of 0.49 for otherwise flared gas, and of 5.55 for otherwise vented gas (all in $CO_2$ eq g / kWh).

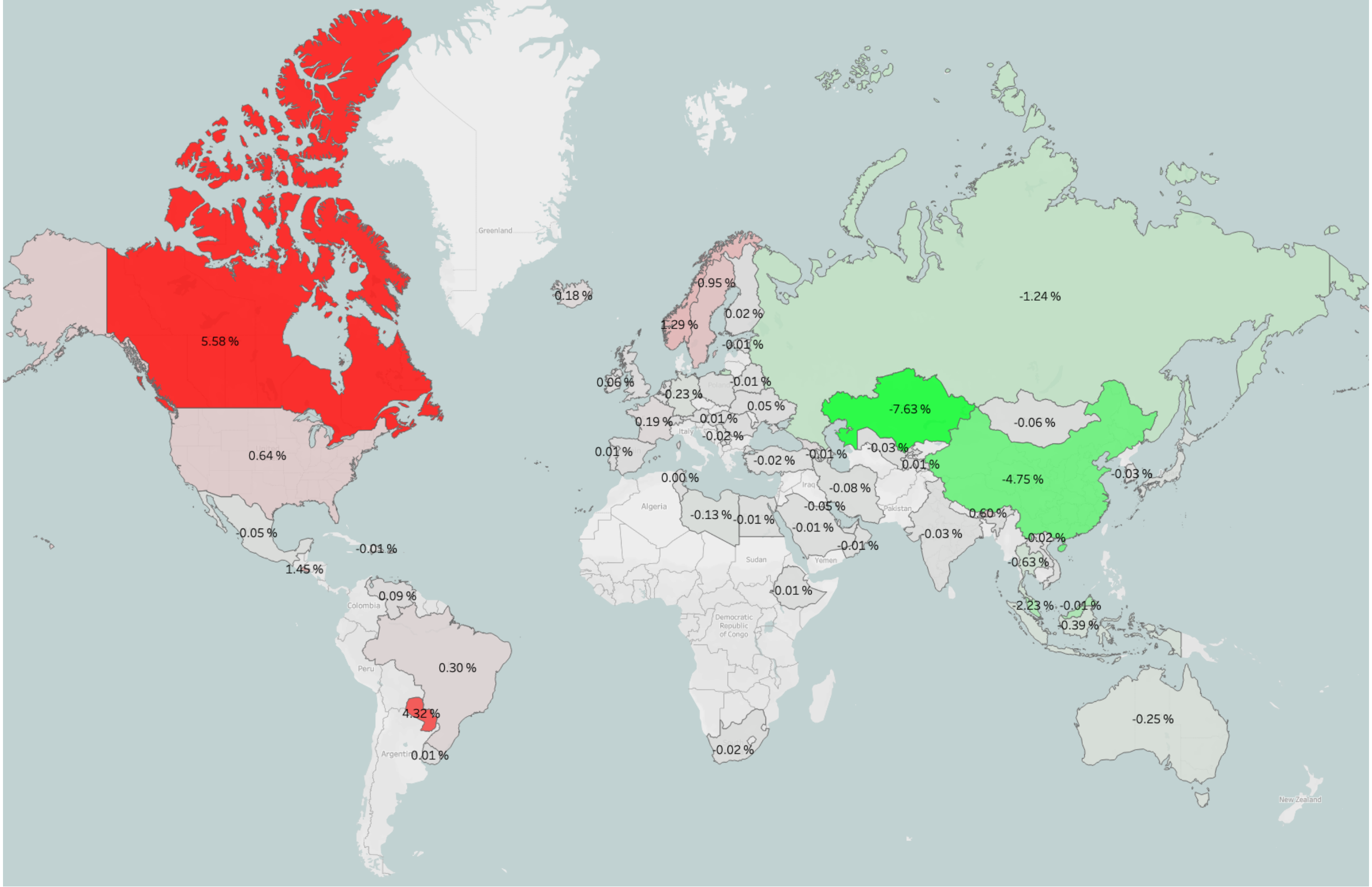

Figure 1. Effect of a Bitcoin mining ban per country as a percentage of global annual emissions from the network. In red, countries where the ban would backfire, resulting in an increase in global POG emissions. In green, where the ban would result in a successful reduction in global emissions.

million tonnes) and Malaysia (-997,000 tonnes) would similarly see reductions in their overall carbon emissions following a ban.

At the state level, results in the United States vary significantly. States like Kentucky (-2.65 million tonnes), Georgia (-632,000 tonnes), and Nebraska (-272,000 tonnes) would reduce POG emissions if a ban were imposed. However, states with greener grids and substantial off-grid mining operations, such as New York (+1.5 million tonnes) and Texas (+1 million tonnes), would experience increased POG emissions. As a result, a ban in the US would increase the total emissions by 287,000 tonnes.

Similarly for China, results vary significantly. A ban in Xinjiang is projected to decrease carbon emissions by 3.1 million tonnes. However, for regions with greener grids like Sichuan and Yunnan, the POG emissions would increase by 1.7 million and 415,000 tonnes of $CO_2$ respectively.

These results are all roughly cut in halfiflimited ban effectiveness is assumed.

## IV. Discussion

### A. Implications

The findings reveal a significant paradox in the environmental regulation of Bitcoin mining. While the primary intent behind imposing mining bans in low-carbon jurisdictions is to curb carbon emissions, our results demonstrate that such policies can have the unintended consequence ofincreasing them. This occurs as a result of the the relocation of mining activities to regions where electricity generation is more carbon-intensive than their former jurisdiction. As a result, while a ban might reduce emissions locally, it can lead to a net increase in global emissions, undermining the original (global) environmental goals.

This suggests that policymakers should reconsider the effectiveness of outright Bitcoin mining bans as a tool for reducing global carbon emissions, which could be counter-productive. Instead, alternative regulatory approaches could prove more effective, such as incentivizing the use of renewable energy for mining operations in high-carbon jurisdiction, and attracting bitcoin mining operations to low-carbon jurisdictions through tax breaks and other benefits. These measures could be more aligned with the broader goals of reducing global

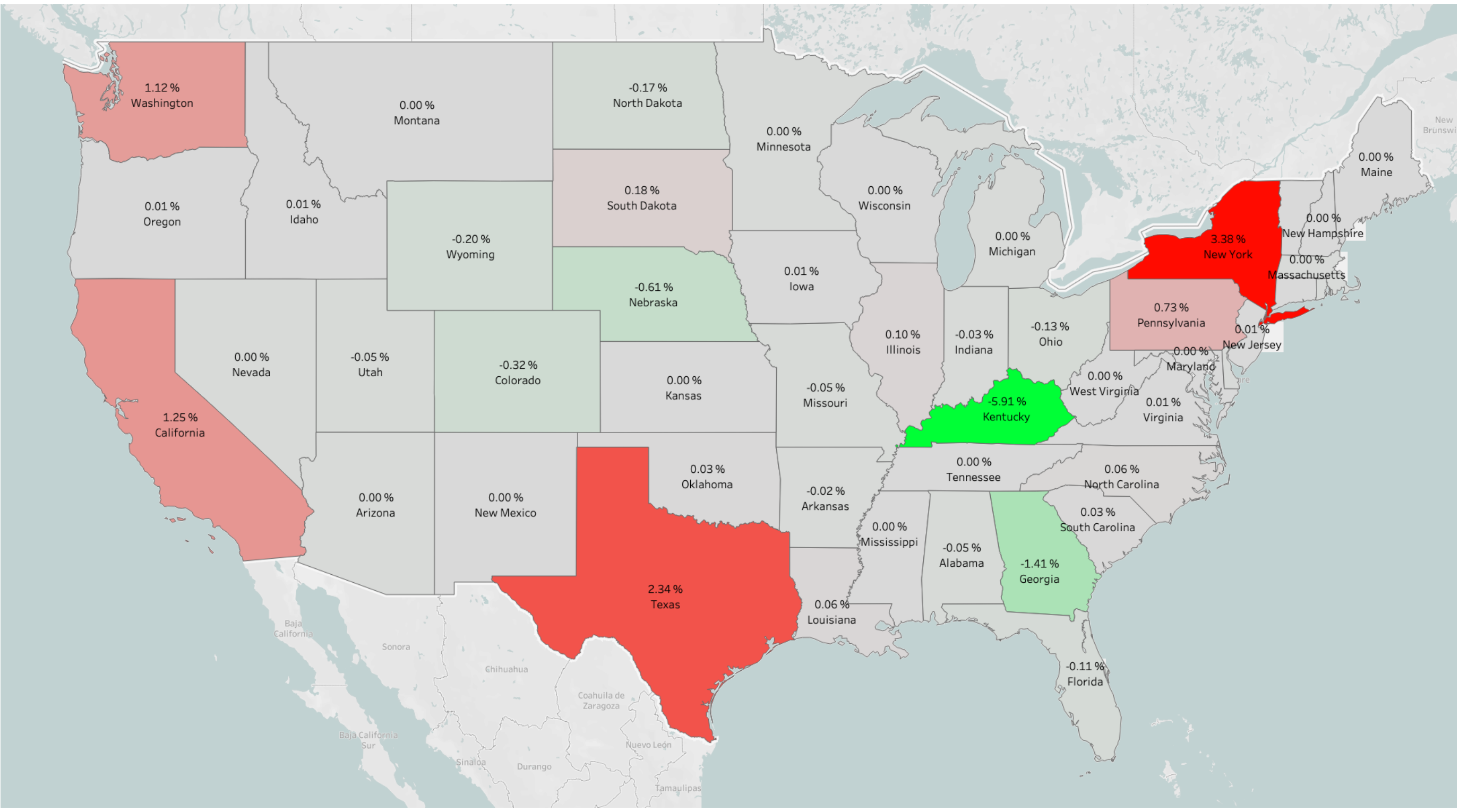

Figure 2. Effect of a Bitcoin mining ban key US jurisdictions as a percentage of global annual emissions from the network. In red, states where the ban would backfire, resulting in an increase in global POG emissions. In green, where the ban would result in a successful reduction in global emissions.

carbon emissions while allowing the cryptocurrency industry to continue its growth.

From a methodological point of view, this study shows the continued importance of the current proportion hash rate redistribution method, which should be used together with the driving forces method to estimate a range of effects. Future research should explore further the combination of both methods and the introduction of additional refinements in the driving force one. Limitations should be nevertheless noted, notably the uncertainty on hash rate location in spite of the efforts made to update and validate the data. Future research should also seek to update the state of the art on miner location identification, particularly concerning off-grid activities. In addition, the reader should note that the estimated effect for countries with low hash rate provides very limited insights into policy effectiveness, as country-level results may be biased by the influence of one or two large sites. Finally, we must consider that, to the extent that the driving-force method holds explanatory power, most of the effects identified in this article would broadly be expected to be mitigated in part: Low and high carbon jurisdiction display a degree of geographical clustering, meaning that insofar miners move to nearby jurisdictions instead of globally redistributing themselves, effects would be smaller as the destination country could be characterized by a similar carbon intensity of the country of origin.

Additionally, further research into alternative policy measures is warranted. Exploring the potential ofincentive-based approaches, such as tax incentives for renewable energy use or for relocation towards low-carbon jurisdictions, could provide valuable insights into how best to regulate the environmental impact of cryptocurrency mining. Finally, longitudinal studies that track the long-term effects of mining bans and the evolving carbon footprint of the Bitcoin network would be beneficial. Overall, this stresses the importance of science-based approaches to policymaking .

## B. *Carbon leakage and policy externalities*

These findings broadly are consistent with multiple other corpora ofliterature under the overarching genre of policy externality studies. The most well-known example of policy externality is carbon leakage, a situation by which climate policies in pursuit of reducing domestic emissions end up pushing carbon-intensive manufacturing to less stringent countries, such that global emissions do not fall (or do not fall as much as expected) as a result but the offshoring country could claim credit for an exaggerated apparent domestic reduction. This phenomenon is considered especially egregious when production is displaced, but the consumption that incentivises this production still takes place at the displacing country, and has therefore led researchers to call for consumption-based carbon accounting [33].

In this direction, see Call for tenders FISMA/2023/OP/0005.

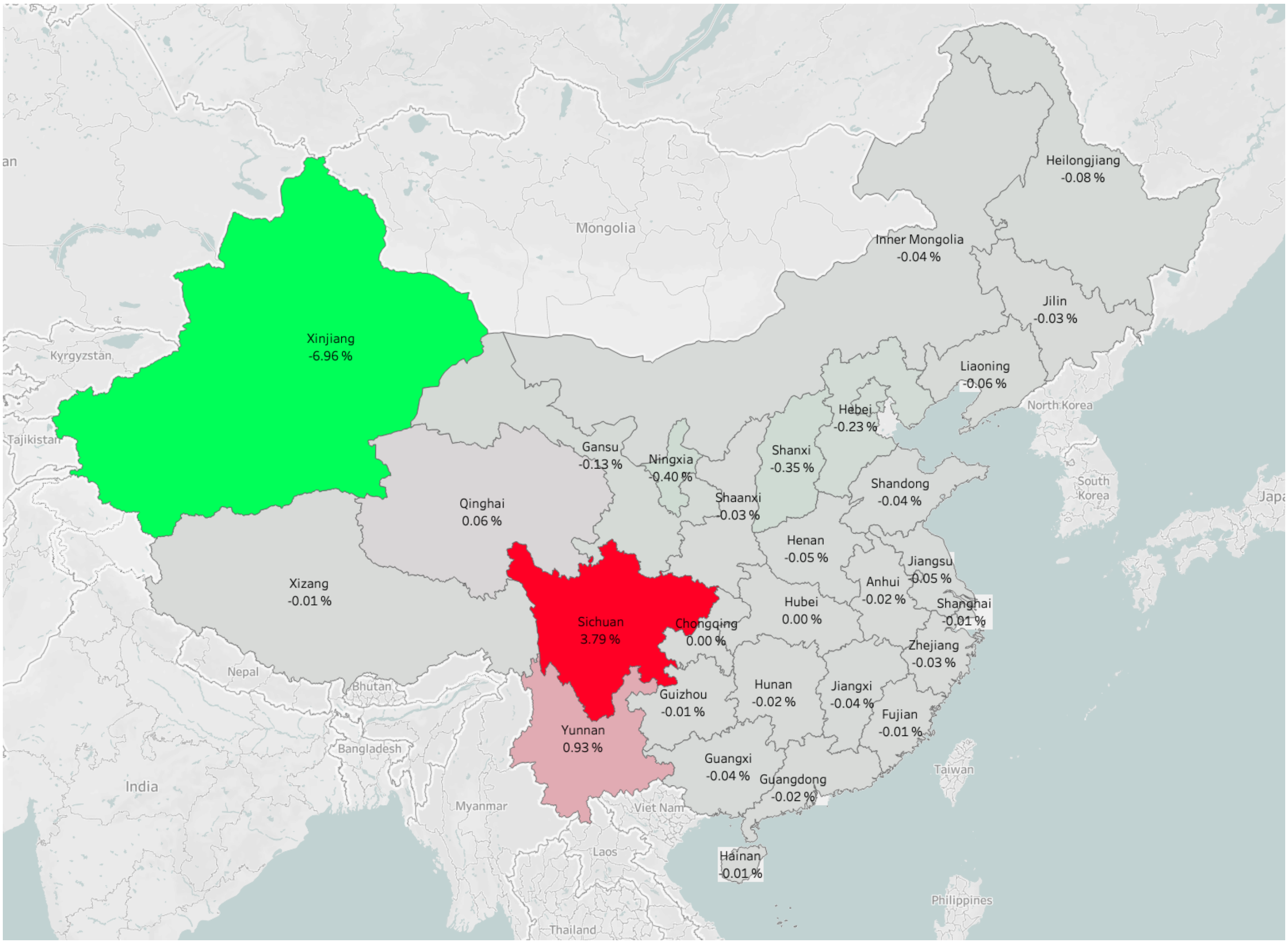

Figure 3. Effect of a Bitcoin mining ban in select Chinese provinces as a percentage of global annual emissions from the network. In red, provinces where the ban would backfire, resulting in an increase in global POG emissions. In green, where the ban would result in a successful reduction in global emissions.

Carbon leakage thus occurs when domestic climate policies result not in a reduction of global emissions but in an increase ofimported emissions relative to domestic emissions [33], [34]. Furthermore, carbon leakage can entail an increase in global emissions when the "carbon leakage rate" is greater than 1, that is, for every ton of $CO_2$ reduced locally, more than a ton begins to be emitted outside of the jurisdiction. [35] Making matters worse, carbon leakage is also associated with deadweight losses, making the policy intervention both ineffective and inefficient. [35]

Resort to the carbon leakage literature thus constitutes a theoretical frame of reference of high significance for this study's inquiry. For instance, [36] identify three preconditions for high leakage risk:

1) That the regulated sector faces aggressive rivals from abroad.
2) That the regulated product is homogeneous such that customers are unable to distinguish between goods made within the regulated jurisdiction and outside ofit.
3) That consumers are price sensitive, which makes harder for firms to pass costs through to them.

It is clear that the first two preconditions are clearly met. The design of Bitcoin is such that it has led to an almost perfectly competitive market [37] with highly aggressive competition within and without national frontiers. The bitcoin asset is almost perfectly fungible, with consumers being partially able but in practice not willing to distinguish coins from one another based on their origin [38], [39], especially for climatic reasons. Bitcoin buyers are not exactly price sensitive. Although bitcoin is not regarded as a Veblen good [40], it is generally accepted that bitcoin demand can often increase with price and that segments ofit actually do exhibit Veblen effects, which is the opposite of price elasticity [41]. Nevertheless, for this condition to be relevant, price pass-through should be possible for producers via supply restrictions, but this is not the case, as bitcoin supply is algorithmically determined and a reduction in the amount of hashing does not result in a fall in the amount of bitcoin mined. It is really the ability for price-increase pass-through that matters, not the price elasticity in itself.

One should also note that carbon leakage sometimes has been regarded as a problem of relatively minor importance because energy costs account for a small percentage of total costs [42]. However, the reverse is the cost for the Bitcoin mining market, where energy costs constitute the around half of the producer's expenses [31]. This implies that carbon leakage can be especially problematic in this particular market.

In consequence, the phenomenon identified in this article can be clearly defined as carbon leakage, with a number of cases of "worst forms of carbon leakage", whereby the leakage results in an increase in carbon emissions [43]. This calls for additional research in the area, exploring parallelisms between existing best practices to prevent carbon leakage, and Bitcoin regulation policy. Nonetheless, carbon leakage is not the only thinkable form of policy externality. Domestic policies against hazardous and electronic waste (e-waste) can have a similar exporting effect, which could even violate the spirit of the Basel Convention, namely to dump hazardous waste (including the e-waste from bitcoin ASICs) from developed countries to developing countries [44].

## V. Conclusion

Our study highlights a critical paradox in the environmental regulation of Bitcoin mining. While bans on Bitcoin mining in low-carbon regions are intended to reduce carbon emissions, our findings suggest that such policies may inadvertently lead to increased global carbon emissions by shifting mining activities to more carbon-intensive regions. This relocation is a form of carbon leakage that not only may undermine the environmental objectives of the bans but also can exacerbate the overall carbon footprint of Bitcoin mining. Therefore, it is imperative for policymakers to consider nuanced approaches to enhancing the positive environmental impacts of cryptocurrencies and mitigating their negative ones. Our research provides a data-driven foundation for developing such strategies, emphasising the need for data-driven regulatory frameworks.

## Acknowledgements

The authors thank Exponential Science, and Exponential Science Capital's Nodiens for access to technical resources that facilitated this research project.

## Conflict of Interest

The authors declare that they have no known competing financial interests or personal relationships that could have appeared to influence the work reported in this paper.

## Acronyms

**ASIC** application-specific integrated circuit
**BEEST** Bitcoin Energy & Emissions Sustainability Tracker
**CBECI** Cambridge Bitcoin Electricity Consumption Index
**CCAF** Cambridge Centre for Alternative Finance
**$CO_2$** carbon dioxide
**DA-RI** Digital Assets Research Institute
**LCA** life cycle assessment
**POG** point-of-generation
**PoW** proof-of-work
**PUE** Power Usage Effectiveness
**ROW** rest-of-the-world
**SEC** US Securities and Exchange Commission
**VPNs** Virtual Private Networks

# Appendix

| Country | Full Effectiveness | | Limited Effectiveness | |
|---|---|---|---|---|
| | LCA | POG | LCA | POG |
| Canada | 2,389.66 | 2,494.93 | 1,194.83 | 1,247.47 |
| Paraguay | 1,839.86 | 1,931.09 | 919.93 | 965.55 |
| Singapore | 1,007.39 | 975.48 | 503.70 | 487.74 |
| El Salvador | 591.40 | 650.00 | 295.70 | 325.00 |
| Norway | 549.03 | 576.27 | 274.51 | 288.13 |
| European Union | 488.57 | 522.78 | 244.28 | 261.39 |
| Sweden | 410.00 | 426.64 | 205.00 | 213.32 |
| United States | 236.42 | 286.63 | 118.21 | 143.32 |
| Bhutan | 254.81 | 267.45 | 127.40 | 133.72 |
| Brazil | 125.91 | 132.87 | 62.96 | 66.44 |
| France | 84.46 | 85.99 | 42.23 | 42.99 |
| Iceland | 75.89 | 80.44 | 37.95 | 40.22 |
| Georgia | 67.00 | 70.21 | 33.50 | 35.10 |
| Venezuela | 39.24 | 41.30 | 19.62 | 20.65 |
| United Kingdom | 34.29 | 35.07 | 17.14 | 17.54 |
| Ireland | 37.61 | 27.62 | 18.81 | 13.81 |
| Ukraine | 21.02 | 21.44 | 10.51 | 10.72 |
| Switzerland | 16.42 | 15.92 | 8.21 | 7.96 |
| Romania | 14.91 | 15.49 | 7.45 | 7.74 |
| Netherlands | 9.06 | 10.48 | 4.53 | 5.24 |
| Finland | 8.61 | 8.87 | 4.30 | 4.43 |
| Spain | 6.01 | 6.39 | 3.00 | 3.19 |
| Tajikistan | 5.48 | 5.78 | 2.74 | 2.89 |
| Hungary | 5.47 | 5.77 | 2.74 | 2.89 |
| Uruguay | 3.81 | 4.00 | 1.90 | 2.00 |
| Belgium | 3.84 | 3.93 | 1.92 | 1.96 |
| Laos | 2.64 | 2.93 | 1.32 | 1.46 |
| Austria | 2.55 | 2.69 | 1.27 | 1.35 |
| Portugal | 2.55 | 2.67 | 1.27 | 1.34 |
| Albania | 1.79 | 1.88 | 0.89 | 0.94 |
| Angola | 1.53 | 1.62 | 0.76 | 0.81 |
| Colombia | 1.33 | 1.43 | 0.67 | 0.72 |
| Croatia | 1.09 | 1.15 | 0.54 | 0.57 |
| Armenia | 1.08 | 1.09 | 0.54 | 0.54 |
| New Zealand | 1.00 | 1.06 | 0.50 | 0.53 |
| Denmark | 0.95 | 0.99 | 0.47 | 0.50 |
| Costa Rica | 0.89 | 0.94 | 0.45 | 0.47 |
| Latvia | 0.86 | 0.91 | 0.43 | 0.45 |
| Luxembourg | 0.84 | 0.89 | 0.42 | 0.45 |
| Greece | 0.27 | 0.59 | 0.14 | 0.29 |
| Bulgaria | 0.42 | 0.46 | 0.21 | 0.23 |
| Ecuador | 0.32 | 0.34 | 0.16 | 0.17 |
| Slovenia | 0.25 | 0.27 | 0.13 | 0.13 |
| Guatemala | 0.14 | 0.18 | 0.07 | 0.09 |
| Kyrgyzstan | 0.14 | 0.15 | 0.07 | 0.07 |
| Panama | 0.13 | 0.13 | 0.06 | 0.07 |
| Lithuania | 0.13 | 0.13 | 0.06 | 0.07 |
| Afghanistan | 0.11 | 0.12 | 0.06 | 0.06 |
| Cameroon | 0.09 | 0.09 | 0.04 | 0.05 |
| Kenya | 0.06 | 0.06 | 0.03 | 0.03 |
| Slovakia | 0.06 | 0.06 | 0.03 | 0.03 |
| Myanmar | 0.05 | 0.05 | 0.02 | 0.03 |
| Honduras | 0.04 | 0.05 | 0.02 | 0.02 |
| Monaco | 0.04 | 0.04 | 0.02 | 0.02 |
| Sudan | 0.04 | 0.04 | 0.02 | 0.02 |
| Peru | 0.02 | 0.02 | 0.01 | 0.01 |
| Zimbabwe | 0.01 | 0.01 | 0.00 | 0.00 |
| Bahamas | -0.01 | -0.01 | 0.00 | 0.00 |
| Pakistan | -0.01 | -0.01 | -0.01 | -0.01 |
| Ghana | -0.02 | -0.02 | -0.01 | -0.01 |
| Cambodia | -0.03 | -0.03 | -0.02 | -0.01 |
| Puerto Rico | -0.03 | -0.03 | -0.02 | -0.02 |
| Chile | -0.06 | -0.05 | -0.03 | -0.02 |
| Qatar | -0.07 | -0.07 | -0.03 | -0.04 |
| Bolivia | -0.08 | -0.09 | -0.04 | -0.04 |
| Trinidad and Tobago | -0.08 | -0.09 | -0.04 | -0.04 |
| Moldova | -0.09 | -0.09 | -0.04 | -0.05 |
| Algeria | -0.12 | -0.13 | -0.06 | -0.06 |
| Philippines | -0.14 | -0.14 | -0.07 | -0.07 |
| Israel | -0.20 | -0.20 | -0.10 | -0.10 |
| Sri Lanka | -0.21 | -0.21 | -0.10 | -0.10 |
| Nigeria | -0.27 | -0.29 | -0.14 | -0.14 |
| Turkmenistan | -0.31 | -0.33 | -0.15 | -0.16 |
| Iraq | -0.38 | -0.39 | -0.19 | -0.20 |
| Maldives | -0.40 | -0.40 | -0.20 | -0.20 |
| Czechia | -0.41 | -0.40 | -0.20 | -0.20 |
| Cyprus | -0.42 | -0.42 | -0.21 | -0.21 |
| Bosnia and Herzegovina | -0.51 | -0.51 | -0.26 | -0.26 |
| Seychelles | -0.61 | -0.62 | -0.31 | -0.31 |
| Italy | -1.32 | -0.67 | -0.66 | -0.33 |
| Lebanon | -0.70 | -0.71 | -0.35 | -0.36 |
| Morocco | -0.86 | -0.87 | -0.43 | -0.44 |
| Taiwan | -0.91 | -0.93 | -0.46 | -0.47 |
| North Macedonia | -1.18 | -1.19 | -0.59 | -0.59 |
| Argentina | -1.25 | -1.36 | -0.62 | -0.68 |
| United Arab Emirates | -1.48 | -1.56 | -0.74 | -0.78 |
| Bahrain | -1.59 | -1.68 | -0.79 | -0.84 |
| Bangladesh | -1.76 | -1.82 | -0.88 | -0.91 |
| Azerbaijan | -2.12 | -2.24 | -1.06 | -1.12 |
| Saudi Arabia | -2.47 | -2.55 | -1.23 | -1.27 |
| Dominican Republic | -2.83 | -2.91 | -1.42 | -1.45 |
| Belarus | -3.77 | -4.01 | -1.89 | -2.00 |
| Oman | -3.88 | -4.08 | -1.94 | -2.04 |
| Brunei | -3.94 | -4.10 | -1.97 | -2.05 |
| Egypt | -4.04 | -4.23 | -2.02 | -2.11 |
| Estonia | -4.71 | -4.45 | -2.35 | -2.22 |
| Ethiopia | -5.13 | -5.26 | -2.56 | -2.63 |
| Turkey | -6.97 | -6.90 | -3.49 | -3.45 |
| Kosovo | -7.59 | -7.72 | -3.79 | -3.86 |
| South Africa | -7.81 | -7.92 | -3.90 | -3.96 |
| Poland | -8.91 | -8.99 | -4.45 | -4.50 |
| Vietnam | -9.51 | -9.35 | -4.75 | -4.68 |
| Serbia | -10.21 | -10.28 | -5.11 | -5.14 |
| South Korea | -13.32 | -13.63 | -6.66 | -6.82 |
| Uzbekistan | -13.53 | -14.21 | -6.77 | -7.11 |
| India | -14.38 | -14.54 | -7.19 | -7.27 |
| Mexico | -21.54 | -22.19 | -10.77 | -11.09 |
| Kuwait | -22.52 | -23.33 | -11.26 | -11.66 |
| Mongolia | -27.87 | -28.38 | -13.93 | -14.19 |
| Iran | -33.66 | -35.39 | -16.83 | -17.69 |
| Japan | -54.87 | -54.85 | -27.43 | -27.42 |
| Libya | -54.95 | -57.18 | -27.48 | -28.59 |
| Germany | -121.16 | -101.76 | -60.58 | -50.88 |
| Australia | -110.64 | -110.46 | -55.32 | -55.23 |
| Indonesia | -172.26 | -174.27 | -86.13 | -87.13 |
| Hong Kong | -216.41 | -223.11 | -108.20 | -111.55 |
| Thailand | -274.70 | -281.81 | -137.35 | -140.90 |
| Russia | -526.57 | -552.99 | -263.28 | -276.50 |
| Malaysia | -977.26 | -997.11 | -488.63 | -498.55 |
| China | -1,894.57 | -2,126.51 | -947.28 | -1,063.26 |
| Kazakhstan | -3,345.54 | -3,411.55 | -1,672.77 | -1,705.78 |

Table I: Global effect of a Bitcoin mining ban, in thousands of carbon dioxide equivalent tonnes per year.

| State | Full Effectiveness | | Limited Effectiveness | |
|---|---|---|---|---|
| | LCA | POG | LCA | POG |
| New York | 1,446.58 | 1,513.70 | 723.29 | 756.85 |
| Texas | 1,059.48 | 1,044.81 | 529.74 | 522.40 |
| California | 501.32 | 556.93 | 250.66 | 278.47 |
| Washington | 480.29 | 501.30 | 240.15 | 250.65 |
| Pennsylvania | 317.49 | 327.44 | 158.74 | 163.72 |
| South Dakota | 81.55 | 82.39 | 40.78 | 41.20 |
| Illinois | 44.58 | 44.20 | 22.29 | 22.10 |
| Louisiana | 27.50 | 26.64 | 13.75 | 13.32 |
| North Carolina | 24.41 | 24.85 | 12.20 | 12.43 |
| Oklahoma | 20.99 | 14.97 | 10.50 | 7.48 |
| South Carolina | 13.97 | 13.87 | 6.98 | 6.94 |
| Virginia | 6.98 | 6.26 | 3.49 | 3.13 |
| Idaho | 5.95 | 6.25 | 2.98 | 3.12 |
| Iowa | 3.54 | 3.35 | 1.77 | 1.67 |
| Oregon | 2.79 | 2.90 | 1.39 | 1.45 |
| New Jersey | 2.43 | 2.39 | 1.21 | 1.20 |
| Tennessee | 2.18 | 2.15 | 1.09 | 1.08 |
| Maine | 0.83 | 0.88 | 0.42 | 0.44 |
| Maryland | 0.71 | 0.71 | 0.35 | 0.35 |
| Wisconsin | 0.17 | 0.24 | 0.08 | 0.12 |
| Kansas | 0.23 | 0.22 | 0.12 | 0.11 |
| Connecticut | 0.09 | 0.01 | 0.05 | 0.01 |
| Arizona | -0.26 | -0.26 | -0.13 | -0.13 |
| Massachusetts | -0.87 | -0.79 | -0.43 | -0.39 |
| Michigan | -1.08 | -1.14 | -0.54 | -0.57 |
| Minnesota | -0.03 | -1.30 | -0.01 | -0.65 |
| Other | -1.73 | -1.81 | -0.86 | -0.90 |
| Nevada | -3.26 | -1.91 | -1.63 | -0.95 |
| Rhode Island | -3.28 | -3.37 | -1.64 | -1.69 |
| Arkansas | -6.86 | -7.11 | -3.43 | -3.56 |
| Delaware | -10.97 | -11.40 | -5.48 | -5.70 |
| Indiana | -11.63 | -11.94 | -5.82 | -5.97 |
| Alabama | -18.12 | -20.56 | -9.06 | -10.28 |
| Missouri | -21.40 | -21.91 | -10.70 | -10.95 |
| Utah | -23.00 | -23.20 | -11.50 | -11.60 |
| Florida | -49.08 | -51.23 | -24.54 | -25.62 |
| Ohio | -53.89 | -56.03 | -26.94 | -28.02 |
| North Dakota | -75.88 | -78.04 | -37.94 | -39.02 |
| Wyoming | -88.01 | -89.97 | -44.00 | -44.99 |
| Colorado | -141.48 | -142.10 | -70.74 | -71.05 |
| Nebraska | -261.53 | -272.26 | -130.76 | -136.13 |
| Georgia | -604.42 | -632.66 | -302.21 | -316.33 |
| Kentucky | -2,591.56 | -2,644.78 | -1,295.78 | -1,322.39 |

Table II: US State-level effect of a Bitcoin mining ban, in thousands of carbon dioxide equivalent tonnes per year.

| State | Full Effectiveness | | Limited Effectiveness | |
|---|---|---|---|---|
| | LCA | POG | LCA | POG |
| Sichuan | 1,786.95 | 1,696.29 | 893.48 | 848.14 |
| Yunnan | 441.84 | 415.26 | 220.92 | 207.63 |
| Qinghai | 28.80 | 27.47 | 14.40 | 13.74 |
| Hubei | -0.39 | -0.85 | -0.20 | -0.43 |
| Chongqing | -1.95 | -2.11 | -0.97 | -1.05 |
| Hainan | -2.44 | -2.60 | -1.22 | -1.30 |
| Xizang | -4.45 | -4.47 | -2.22 | -2.24 |
| Shanghai | -4.42 | -4.60 | -2.21 | -2.30 |
| Fujian | -5.51 | -5.81 | -2.75 | -2.90 |
| Guizhou | -6.02 | -6.57 | -3.01 | -3.28 |
| Hunan | -6.61 | -6.97 | -3.31 | -3.49 |
| Other | -10.23 | -10.29 | -5.11 | -5.14 |
| Anhui | -10.16 | -10.38 | -5.08 | -5.19 |
| Guangdong | -9.97 | -10.67 | -4.99 | -5.34 |
| Jilin | -11.97 | -12.19 | -5.98 | -6.09 |
| Zhejiang | -11.69 | -12.21 | -5.85 | -6.11 |
| Shaanxi | -12.17 | -12.53 | -6.09 | -6.27 |
| Guangxi | -17.09 | -17.85 | -8.55 | -8.93 |
| Shandong | -17.64 | -18.05 | -8.82 | -9.02 |
| Inner Mongolia | -18.43 | -18.69 | -9.21 | -9.34 |
| Jiangxi | -19.05 | -19.66 | -9.53 | -9.83 |
| Henan | -23.82 | -24.36 | -11.91 | -12.18 |
| Jiangsu | -24.00 | -24.60 | -12.00 | -12.30 |
| Liaoning | -28.35 | -28.81 | -14.17 | -14.40 |
| Heilongjiang | -35.93 | -36.63 | -17.97 | -18.32 |
| Gansu | -53.58 | -57.09 | -26.79 | -28.54 |
| Hebei | -99.45 | -100.70 | -49.73 | -50.35 |
| Beijing | -101.17 | -104.36 | -50.58 | -52.18 |
| Shanxi | -152.60 | -155.41 | -76.30 | -77.71 |
| Ningxia | -176.28 | -179.35 | -88.14 | -89.67 |
| Xinjiang | -3,047.42 | -3,115.04 | -1,523.71 | -1,557.52 |

Table III: China State-level effect of a Bitcoin mining ban, in thousands of carbon dioxide equivalent tonnes per year.

| Country | State | One-off Effect Full Effectiveness | | One-off Effect Limited Effectiveness | |
|---|---|---|---|---|---|
| | | LCA | POG | LCA | POG |
| United States | Georgia | 539.67 | 526.25 | 269.83 | 263.12 |
| Kazakhstan | - | 497.48 | 495.29 | 248.74 | 247.64 |
| China | Xinjiang | 415.06 | 415.06 | 207.53 | 207.53 |
| United States | Kentucky | 382.12 | 380.86 | 191.06 | 190.43 |
| Russia | - | 240.14 | 236.03 | 120.07 | 118.01 |
| Malaysia | - | 186.01 | 184.29 | 93.01 | 92.14 |
| Germany | - | 140.10 | 134.47 | 70.05 | 67.24 |
| United States | Texas | 104.68 | 99.96 | 52.34 | 49.98 |
| United States | Nebraska | 96.58 | 95.11 | 48.29 | 47.56 |
| China | Sichuan | 87.70 | 87.70 | 43.85 | 43.85 |
| Canada | - | 96.38 | 79.48 | 48.19 | 39.74 |
| Ireland | - | 80.79 | 78.98 | 40.39 | 39.49 |
| United States | California | 87.33 | 78.87 | 43.67 | 39.43 |
| United States | North Carolina | 74.35 | 71.93 | 37.18 | 35.97 |
| Thailand | - | 63.40 | 62.71 | 31.70 | 31.36 |
| United States | Colorado | 52.36 | 51.14 | 26.18 | 25.57 |
| United States | New York | 52.80 | 42.21 | 26.40 | 21.11 |
| Hong Kong | - | 36.26 | 36.25 | 18.13 | 18.12 |
| China | Yunnan | 33.54 | 33.54 | 16.77 | 16.77 |
| United States | Oklahoma | 33.70 | 33.09 | 16.85 | 16.55 |
| Indonesia | - | 29.38 | 29.08 | 14.69 | 14.54 |
| Australia | - | 24.43 | 23.94 | 12.22 | 11.97 |
| United States | Washington | 27.53 | 23.72 | 13.76 | 11.86 |
| China | Ningxia | 22.84 | 22.84 | 11.42 | 11.42 |
| United States | Alabama | 23.25 | 22.77 | 11.62 | 11.38 |
| United States | Minnesota | 22.55 | 21.95 | 11.28 | 10.98 |
| China | Shanxi | 20.18 | 20.18 | 10.09 | 10.09 |
| United States | North Dakota | 17.74 | 17.56 | 8.87 | 8.78 |
| China | Beijing | 16.92 | 16.92 | 8.46 | 8.46 |
| United States | Florida | 16.32 | 16.11 | 8.16 | 8.06 |
| United States | Wyoming | 14.73 | 14.66 | 7.36 | 7.33 |
| Japan | - | 14.19 | 13.89 | 7.10 | 6.95 |
| China | Gansu | 13.81 | 13.81 | 6.91 | 6.91 |
| United States | Virginia | 12.79 | 12.43 | 6.39 | 6.22 |
| United States | Ohio | 12.21 | 12.15 | 6.11 | 6.07 |
| China | Hebei | 11.60 | 11.60 | 5.80 | 5.80 |
| Libya | - | 10.53 | 10.53 | 5.27 | 5.27 |
| Iran | - | 8.11 | 8.09 | 4.05 | 4.04 |
| Netherlands | - | 8.10 | 7.70 | 4.05 | 3.85 |
| United States | Illinois | 7.89 | 7.56 | 3.95 | 3.78 |
| United Kingdom | - | 6.94 | 6.57 | 3.47 | 3.28 |
| Mexico | - | 6.36 | 6.27 | 3.18 | 3.14 |
| United States | South Dakota | 6.76 | 6.27 | 3.38 | 3.13 |
| China | Heilongjiang | 4.86 | 4.86 | 2.43 | 2.43 |
| United States | Nevada | 5.10 | 4.84 | 2.55 | 2.42 |
| Italy | - | 4.77 | 4.57 | 2.39 | 2.29 |
| United States | Utah | 4.50 | 4.43 | 2.25 | 2.22 |
| Ukraine | - | 4.63 | 4.39 | 2.31 | 2.20 |
| United States | South Carolina | 4.47 | 4.31 | 2.24 | 2.15 |
| South Korea | - | 4.28 | 4.21 | 2.14 | 2.11 |
| Kuwait | - | 4.05 | 4.05 | 2.02 | 2.02 |
| Mongolia | - | 4.06 | 4.04 | 2.03 | 2.02 |
| United States | Missouri | 4.04 | 4.01 | 2.02 | 2.00 |
| Georgia | - | 4.28 | 3.70 | 2.14 | 1.85 |
| China | Jiangsu | 3.60 | 3.60 | 1.80 | 1.80 |
| China | Liaoning | 3.59 | 3.59 | 1.80 | 1.80 |
| China | Guangxi | 3.46 | 3.46 | 1.73 | 1.73 |
| China | Henan | 3.43 | 3.43 | 1.71 | 1.71 |
| Uzbekistan | - | 3.31 | 3.29 | 1.65 | 1.65 |
| China | Jiangxi | 3.19 | 3.19 | 1.60 | 1.60 |
| United States | Delaware | 2.85 | 2.82 | 1.42 | 1.41 |
| China | Guangdong | 2.70 | 2.70 | 1.35 | 1.35 |
| Brazil | - | 3.56 | 2.55 | 1.78 | 1.27 |
| China | Shandong | 2.54 | 2.54 | 1.27 | 1.27 |
| Argentina | - | 2.57 | 2.50 | 1.29 | 1.25 |
| Vietnam | - | 2.57 | 2.50 | 1.28 | 1.25 |
| Venezuela | - | 2.73 | 2.37 | 1.36 | 1.18 |
| China | Zhejiang | 2.36 | 2.36 | 1.18 | 1.18 |
| Romania | - | 2.50 | 2.33 | 1.25 | 1.17 |
| India | - | 2.35 | 2.32 | 1.17 | 1.16 |
| Turkey | - | 2.29 | 2.23 | 1.15 | 1.12 |
| China | Inner Mongolia | 2.23 | 2.23 | 1.12 | 1.12 |
| United States | Arkansas | 2.18 | 2.15 | 1.09 | 1.08 |
| United States | Indiana | 2.10 | 2.09 | 1.05 | 1.04 |
| Serbia | - | 2.12 | 2.09 | 1.06 | 1.04 |
| China | Other | 2.08 | 2.05 | 1.04 | 1.02 |
| China | Shaanxi | 1.98 | 1.98 | 0.99 | 0.99 |
| China | Guizhou | 1.95 | 1.95 | 0.97 | 0.97 |
| United States | Iowa | 1.96 | 1.91 | 0.98 | 0.95 |
| Greece | - | 1.94 | 1.85 | 0.97 | 0.92 |
| United States | Pennsylvania | 3.48 | 1.73 | 1.74 | 0.86 |
| Estonia | - | 1.73 | 1.66 | 0.86 | 0.83 |
| Poland | - | 1.63 | 1.61 | 0.82 | 0.80 |
| China | Jilin | 1.59 | 1.59 | 0.79 | 0.79 |
| China | Hunan | 1.51 | 1.51 | 0.76 | 0.76 |
| China | Anhui | 1.43 | 1.43 | 0.72 | 0.72 |
| Singapore | - | 1.32 | 1.32 | 0.66 | 0.66 |
| France | - | 1.71 | 1.31 | 0.85 | 0.65 |
| China | Hubei | 1.26 | 1.26 | 0.63 | 0.63 |
| China | Fujian | 1.26 | 1.26 | 0.63 | 0.63 |
| United States | Other | 1.27 | 1.24 | 0.64 | 0.62 |
| South Africa | - | 1.19 | 1.18 | 0.60 | 0.59 |
| Belarus | - | 1.16 | 1.16 | 0.58 | 0.58 |
| China | Qinghai | 1.13 | 1.13 | 0.57 | 0.57 |
| United States | Arizona | 1.15 | 1.11 | 0.57 | 0.56 |
| Kosovo | - | 1.11 | 1.10 | 0.55 | 0.55 |
| Egypt | - | 1.08 | 1.07 | 0.54 | 0.54 |
| United States | Tennessee | 1.11 | 1.07 | 0.55 | 0.53 |
| United States | Rhode Island | 1.08 | 1.06 | 0.54 | 0.53 |
| United States | New Jersey | 0.93 | 0.89 | 0.46 | 0.45 |
| China | Xizang | 0.91 | 0.89 | 0.45 | 0.45 |
| Oman | - | 0.87 | 0.87 | 0.44 | 0.44 |
| China | Shanghai | 0.85 | 0.85 | 0.42 | 0.42 |
| United States | Wisconsin | 0.82 | 0.78 | 0.41 | 0.39 |
| Ethiopia | - | 0.77 | 0.77 | 0.39 | 0.39 |
| Sweden | - | 3.28 | 0.75 | 1.64 | 0.37 |
| Brunei | - | 0.74 | 0.74 | 0.37 | 0.37 |
| Laos | - | 0.72 | 0.67 | 0.36 | 0.34 |
| Hungary | - | 0.71 | 0.65 | 0.36 | 0.33 |
| China | Hainan | 0.63 | 0.63 | 0.32 | 0.32 |
| United States | Connecticut | 0.64 | 0.62 | 0.32 | 0.31 |
| China | Chongqing | 0.59 | 0.59 | 0.30 | 0.30 |
| United States | Idaho | 0.63 | 0.57 | 0.32 | 0.29 |
| Azerbaijan | - | 0.55 | 0.54 | 0.27 | 0.27 |
| United States | Massachusetts | 0.56 | 0.53 | 0.28 | 0.27 |
| Dominican Republic | - | 0.53 | 0.53 | 0.27 | 0.27 |
| Spain | - | 0.50 | 0.44 | 0.25 | 0.22 |
| Saudi Arabia | - | 0.42 | 0.42 | 0.21 | 0.21 |
| United States | Michigan | 0.41 | 0.41 | 0.21 | 0.20 |
| United States | Oregon | 0.41 | 0.38 | 0.21 | 0.19 |
| Bahrain | - | 0.36 | 0.36 | 0.18 | 0.18 |
| United Arab Emirates | - | 0.34 | 0.34 | 0.17 | 0.17 |
| Bangladesh | - | 0.33 | 0.33 | 0.16 | 0.16 |
| Belgium | - | 0.31 | 0.29 | 0.16 | 0.14 |
| North Macedonia | - | 0.27 | 0.27 | 0.14 | 0.13 |
| Bulgaria | - | 0.27 | 0.25 | 0.13 | 0.13 |
| United States | Maryland | 0.26 | 0.25 | 0.13 | 0.13 |
| Colombia | - | 0.27 | 0.25 | 0.13 | 0.12 |
| Armenia | - | 0.23 | 0.22 | 0.11 | 0.11 |
| Czechia | - | 0.22 | 0.21 | 0.11 | 0.11 |
| Portugal | - | 0.21 | 0.19 | 0.11 | 0.10 |
| Tajikistan | - | 0.23 | 0.18 | 0.11 | 0.09 |
| Taiwan | - | 0.17 | 0.17 | 0.09 | 0.08 |
| Morocco | - | 0.14 | 0.14 | 0.07 | 0.07 |
| Croatia | - | 0.15 | 0.14 | 0.08 | 0.07 |
| Angola | - | 0.16 | 0.14 | 0.08 | 0.07 |
| United States | Kansas | 0.14 | 0.14 | 0.07 | 0.07 |
| Lebanon | - | 0.11 | 0.11 | 0.06 | 0.06 |
| Finland | - | 0.16 | 0.11 | 0.08 | 0.06 |
| Bosnia and Herzegovina | - | 0.11 | 0.11 | 0.06 | 0.06 |
| Nigeria | - | 0.11 | 0.11 | 0.05 | 0.05 |
| Seychelles | - | 0.11 | 0.10 | 0.05 | 0.05 |
| Uruguay | - | 0.13 | 0.10 | 0.07 | 0.05 |
| Guatemala | - | 0.10 | 0.09 | 0.05 | 0.04 |
| Austria | - | 0.11 | 0.09 | 0.06 | 0.04 |
| Chile | - | 0.09 | 0.08 | 0.04 | 0.04 |
| United States | Maine | 0.09 | 0.08 | 0.05 | 0.04 |
| Cyprus | - | 0.08 | 0.08 | 0.04 | 0.04 |
| Iraq | - | 0.07 | 0.07 | 0.04 | 0.04 |
| Turkmenistan | - | 0.07 | 0.07 | 0.04 | 0.04 |
| Maldives | - | 0.06 | 0.06 | 0.03 | 0.03 |
| Bolivia | - | 0.05 | 0.05 | 0.03 | 0.03 |
| Paraguay | - | 12.30 | 0.05 | 6.15 | 0.02 |
| Latvia | - | 0.05 | 0.05 | 0.03 | 0.02 |
| Sri Lanka | - | 0.05 | 0.05 | 0.02 | 0.02 |
| Israel | - | 0.04 | 0.04 | 0.02 | 0.02 |
| Slovenia | - | 0.04 | 0.04 | 0.02 | 0.02 |
| Denmark | - | 0.04 | 0.04 | 0.02 | 0.02 |
| New Zealand | - | 0.04 | 0.03 | 0.02 | 0.02 |
| Algeria | - | 0.03 | 0.03 | 0.01 | 0.01 |
| Sudan | - | 0.03 | 0.03 | 0.01 | 0.01 |
| Honduras | - | 0.03 | 0.02 | 0.01 | 0.01 |
| Moldova | - | 0.02 | 0.02 | 0.01 | 0.01 |
| Philippines | - | 0.02 | 0.02 | 0.01 | 0.01 |
| Ecuador | - | 0.02 | 0.02 | 0.01 | 0.01 |
| Trinidad and Tobago | - | 0.02 | 0.02 | 0.01 | 0.01 |
| Myanmar | - | 0.02 | 0.02 | 0.01 | 0.01 |
| Ghana | - | 0.02 | 0.02 | 0.01 | 0.01 |
| Macao | - | 0.02 | 0.02 | 0.01 | 0.01 |
| Qatar | - | 0.02 | 0.02 | 0.01 | 0.01 |
| Cameroon | - | 0.02 | 0.02 | 0.01 | 0.01 |
| Luxembourg | - | 0.02 | 0.01 | 0.01 | 0.01 |
| Cambodia | - | 0.01 | 0.01 | 0.01 | 0.01 |
| Lithuania | - | 0.01 | 0.01 | 0.01 | 0.00 |
| Costa Rica | - | 0.02 | 0.01 | 0.01 | 0.00 |
| Panama | - | 0.01 | 0.01 | 0.00 | 0.00 |
| Kyrgyzstan | - | 0.01 | 0.01 | 0.00 | 0.00 |
| Pakistan | - | 0.01 | 0.01 | 0.00 | 0.00 |
| Afghanistan | - | 0.01 | 0.01 | 0.00 | 0.00 |
| Puerto Rico | - | 0.00 | 0.00 | 0.00 | 0.00 |
| Zimbabwe | - | 0.00 | 0.00 | 0.00 | 0.00 |
| Montenegro | - | 0.00 | 0.00 | 0.00 | 0.00 |
| Peru | - | 0.00 | 0.00 | 0.00 | 0.00 |
| Slovakia | - | 0.00 | 0.00 | 0.00 | 0.00 |
| Bahamas | - | 0.00 | 0.00 | 0.00 | 0.00 |
| Switzerland | - | 0.00 | 0.00 | 0.00 | 0.00 |
| Kenya | - | 0.00 | 0.00 | 0.00 | 0.00 |
| Monaco | - | 0.00 | 0.00 | 0.00 | 0.00 |
| Albania | - | 0.01 | 0.00 | 0.01 | 0.00 |
| Benin | - | 0.00 | 0.00 | 0.00 | 0.00 |
| Bhutan | - | 1.76 | 0.00 | 0.88 | 0.00 |
| Bouvet Island | - | 0.00 | 0.00 | 0.00 | 0.00 |
| Congo | - | 0.00 | 0.00 | 0.00 | 0.00 |
| Cuba | - | 0.00 | 0.00 | 0.00 | 0.00 |
| Democratic Republic of Congo | - | 0.00 | 0.00 | 0.00 | 0.00 |
| El Salvador | - | 6.53 | 0.00 | 3.27 | 0.00 |
| Guinea-Bissau | - | 0.00 | 0.00 | 0.00 | 0.00 |
| Iceland | - | 0.59 | 0.00 | 0.30 | 0.00 |
| Madagascar | - | 0.00 | 0.00 | 0.00 | 0.00 |
| Norway | - | 3.77 | 0.00 | 1.88 | 0.00 |
| Tunisia | - | 0.00 | 0.00 | 0.00 | 0.00 |
| Zambia | - | 0.00 | 0.00 | 0.00 | 0.00 |
| United States | Alaska | 0.00 | 0.00 | 0.00 | 0.00 |
| United States | Hawaii | 0.00 | 0.00 | 0.00 | 0.00 |
| United States | Mississippi | 0.00 | 0.00 | 0.00 | 0.00 |
| United States | Montana | 0.00 | 0.00 | 0.00 | 0.00 |
| United States | New Hampshire | 0.00 | 0.00 | 0.00 | 0.00 |
| United States | New Mexico | 0.00 | 0.00 | 0.00 | 0.00 |
| United States | Vermont | 0.00 | 0.00 | 0.00 | 0.00 |
| United States | West Virginia | 0.00 | 0.00 | 0.00 | 0.00 |
| United States | Louisiana | 0.00 | 0.00 | 0.00 | 0.00 |

Table IV: One-off Carbon Avoidance Due to Ban's Frictional Relocation Effects. In Thousands of carbon dioxide equivalent tonnes.

## Exponential Science

Exponential Science integrates exponential technologies to address complex global challenges through education, research, and innovation. Led by Dr. Paolo Tasca and Nikhil Vadgama, Exponential Science collaborates with academia, industry, and government to foster deep tech knowledge, generate pioneering research, and promote early-stage innovations. Exponential Science, as a natural evolution of the DSF's work, aims to harness the convergence of technologies such as blockchain, AI, and IoT to create transformative impacts towards a more decentralized and open society.